\documentclass{ws-ijmpe}

\begin{document}

\markboth{Y. Abe, C. Shen, D. Boilley and B.G. Giraud}{From Di-Nucleus to Mono-Nucleus}

\catchline{}{}{}{}{}

\title{From Di-Nucleus to Mono-Nucleus  \\ 
 - Neck Evolution in Fusion of Massive Systems - }

\author{ YASUHISA ABE}

\address{Research Center of Nuclear Physics, Osaka Univ., 
10-1 Mihogaoka, Ibaraki, Osaka, Japan\\
abey@rcnp.osaka-u.ac.jp}

\author{CAIWAN SHEN}

\address{School of Science, Huzhou Tearchers College,
Huzhou, 313000/Zhejiangone, P. R. China\\
cwshen@hutc.zj.cn}

\author{DAVID BOILLEY}

\address{ GANIL, CEA/DSM-CNRS/IN3P3, BP 55027, F-14076 Caen cedex 5, France \\
and Univ. of Caen, BP 5186, F-14032 Caen cedex, France\\
boilley@ganil.fr}

\author{BERTRAND G. GIRAUD}

\address{Institut Physique Theorique, Direction des Sciences de la Matiere, Centre d'Etudes-Saclay,
Gif-sur-Yvette. F-91191, France\\
bertrand.giraud@cea.fr}

\maketitle

\begin{history}
\received{(received date)}
\revised{(revised date)}
\end{history}

\begin{abstract}
Dynamics of the neck degree of freedom during fusioning process between heavy ions is studied.  Time scales of the three degrees of freedom (the relative distance, the neck and the mass-asymmetry) are studied, showing an early equilibration of the neck. This means that a di-nucleus formed by the incident combination of ions quickly forms a mono-nucleus with a superdeformation during the fusion process and that the other two degrees of freedom have to be solved in a coupled way. 
A brief introduction of Langevin approach and dissipation-fluctuation dynamics is also given and of the application to the synthesis of the superheavy elements.  
\end{abstract}

\section{Motivation : Historical Background}

It is well known that the collective motions in excited nuclei are governed by the potential landscape and dissipation. When the system has to cross a barrier, the fluctuation, associated with 
the dissipation according to the Dissipation-Fluctuation theorem,  play an essential role.
 
For multi-dimensional problems, the Langevin equation appears to  be easier to solve numerically than  the equivalent Fokker-Planck equation also used for the analysis of heavy-ion 
collisions, say, of the fast-fission process\cite{HD}. The Langevin equation was first used in nuclear physics for a dynamical description of the fission process.\cite{ABE1}  Combined with particle emission, 
it was applied to analyse the anomalous multiplicities of pre-scission neutrons,\cite{WADA}
 which, together with the total kinetic energy of fission fragments, supports a strong friction 
of the one-body type, (Wall-and-Window\cite{BLO}) rather than the two-body viscosity.\cite{DAV}
 
More recently, this Langevin formalism has been used to study the fusion of heavy nuclei in order to  propose an explanation of the fusion hindrance\cite{ABE2}  
 which has been experimentally known to exist in fusion of massive systems\cite{SAH}
without theoretical explanation. The DNC 
(Di-Nucleus Configuration) formed by the contact of two ions of the incident channels has 
an extremely large deformation and then is located outside of the conditional saddle point 
in LDM (Liquid Drop Model), as is shown in Fig. 1. The system has then to cross two barriers to fuse. In order to calculate probability of the hindered fusion, we use a Two-Step Model 
and apply it to the synthesis of the superheavy elements.\cite{SHEN,BOU} 

 \begin{figure}[th]
\centerline{\psfig{file=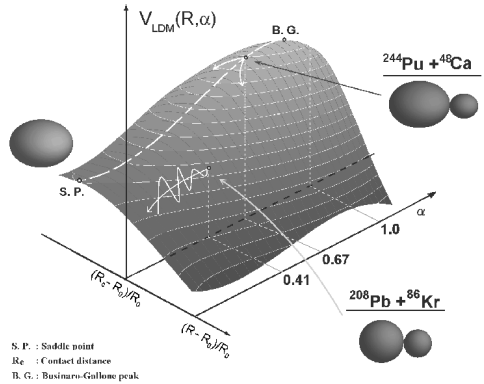,width=8cm}}
\vspace*{5pt}
\caption{A schematic illustration of LDM energy surface for heavy systems.  The x-axis is the distance between two centers and the y-axis is the mass-asymmetry.  Two examples of  incident combinations of ions are indicated for the hot and cold fusion paths, respectively.}
\end{figure}

The over-passing probability of the inner barrier is dominated by the diffusion of the distance parameter between two ions.\cite
{ABE3,ABE4,SWI,BOIL} This is briefly recapitulated in section 2 with a simplified model.  In section 3, time development of the neck motion is analysed with the Smolchowski equation during the fusion process.
  
\section{Brief Reminder of One-Dimensional Parabolic Barrier}

Reducing the inner barrier to an inverted parabola and assuming that the transport coefficients are constant near the barrier, the problem is  amenable to a simple analytic 
explanation of the origin of the fusion hindrance. Here, we will recapitulate one-dimensional case and refer to Ref.\cite{ABE3} for the N-dimensional case with a coupled Langevin equation.   

The equation to be solved is given as follows, 
\begin{equation}
\frac{d}{dt} \left[\begin{array}{c}q \\ p \end{array}\right] = \left[\begin{array}{cc}0&1/\mu\\ \mu\omega^2 & -\beta\end{array}\right].
 \left[\begin{array}{c}q \\ p \end{array}\right] +  \left[\begin{array}{c}0 \\ R\end{array} \right],
\end{equation}  
where $ \mu$ and  $\omega $  denote the inertia mass and the frequency of the parabola, respectively.  
The reduced friction $ \beta = \gamma / \mu $ is defined with the friction coefficient $ \gamma $. 
 R denotes a Langevin force associated to the friction $\gamma$, which is assumed to satisfy 
the dissipation-fluctuation theorem, and to be a Markovian with a Gaussian distribution.   
The probability of passing over the barrier, which we call formation probability of the compound nucleus,  is simply given by an error function, whose 
argument is expressed by the average trajectory $ <q(t)>$ and its variance,
 \begin{eqnarray}
P_{form}(q_0,p_0,t)&=&\int_0^{\infty}\frac{dq}{\sqrt{2\pi}} \frac1{\sigma_q(t)}\exp\left(
-\frac{(q-<q(t)>)^2}{2\sigma^2_q(t)}\right) \\
&=& \frac12 {\rm erfc}\left(-\frac{<q(t)>}{\sqrt2\sigma_q(t)}\right).
\end{eqnarray}
For a time long enough, the probability  converges to a finite value,
\begin{equation}
\lim_{t\to\infty}F_{form}= \frac12 {\rm erfc} \left(\sqrt{\frac{x+ \sqrt{x^2+ 1}}{2x}} \sqrt{\frac{B}T} - \frac1{\sqrt{2x(x+\sqrt{x^2+1}}} \sqrt{\frac{K}T}\right),
\end{equation}
where $ B = \mu\omega^{2} q_{0}^{2}/2 $, the saddle point height measured from the initial point 
$q_{0}$,  while $ K= p_{0}^2/(2 \mu)$.   $ x $ denotes the critical parameter $ \beta /(2\omega) $.  The probability becomes when the initial kinetic energy $K= (x + \sqrt{x^2+ 1})^2 B $, which we call an effective barrier $B_{eff}$ for the case of dissipative dynamics.  This simply explains the origin of the hindrance.  As we discussed elsewhere,\cite{KOS,SHEN} the distribution of $ p_0 $ is expected to be a Boltzmann distribution, and then, an averaging over the initial momentum $p_0$ that is thermally distributed with an average value equal to zero, gives  an extremely simple expression for the formation probability,
\begin{equation} 
\label{Pform}
P_{form}(E_{c.m.})=\frac12 {\rm erfc}\left(\sqrt{\frac{B}T}\right).
\end{equation}
As is clearly seen in Eq. (\ref{Pform}), even if we give a larger incident kinetic energy, the formation and then the fusion probability  increase very slowly through the increase of the temperature of the system, which appears in agreement with the experiment.\cite{SAH}   

Before proceeding to a discussion on the motion of the neck degree of freedom, it is meaningful to have a close look at the time evolution of the fusion.  We analyse time-developments of the trajectory, the formation probability and the current over the saddle by the use of the analytic solution.  The results are shown in the first, the second, and the bottom rows of Fig. 2, respectively. \cite{BOIL}  

\begin{figure}[bth]
\centerline{\psfig{file=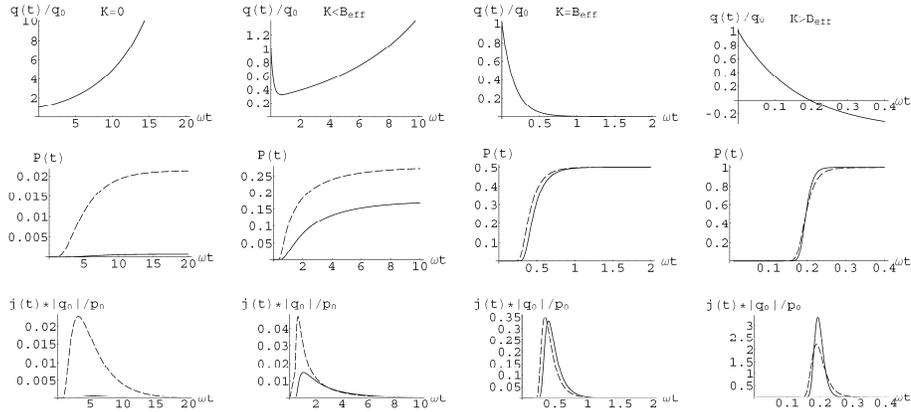,height=5.4cm}}
\caption{Average trajectory, over-passing probability and current at the top of the barrier as a function of time for four regimes, $K=0$ (first column), $K= B_{eff}/2$ (second column), $K= B_{eff} $ (third column) and $K=2 B_{eff}$ (last column). For  each case, two temperatures were chosen, $T=B/5$ (solid line) and $T=B/2$ (dashed line).  Note that each column has a different time scale.}
\end{figure}

Our case corresponds to the leftmost column, for incident kinetic energy $K=0$.  Firstly, the average trajectory never passes over the saddle, but retreats back.  Secondly, the formation probability becomes  saturated as time goes.  Finally, the current has a peak structure around several $\hbar$/MeV, which shows that the probability current at the saddle starts slowly and then terminates gradually.  This means that the formation of the compound nucleus is not due to a dynamical motion, but due to the fluctuation, i.e., due to a tail part of the Gaussian distribution around the average trajectory.  The time scale of the  radial fusion is important for the discussion of the neck degree of freedom in the next section. 

\section{From Di-Nucleus to Mono-Nucleus :  Filling-in of the Neck Cleft}  

For a description of fusion processes between heavy ions, there are at least three parameters or variables.  
In two-center parameterization,\cite{TCP} they are distance between two mass centers, mass-asymmetry, and the neck correction.  Since the neck degree of freedom is weakly coupled to the others, it is meaningful to analyse its time evolution separately.  The LDM potential for the symmetric incident systems turns out to be approximately linear in the neck parameter.  To analyse the time evolution of the neck parameter, starting at $ \epsilon =1.0 $ or around, a Langevin equation is solved. 
It appears that the average value of the neck parameter changes very quickly, far quicker than the radial fusion for most systems including very heavy ones.\cite{NECK}  This is due to the action of the linear driving force in the neck $ \epsilon $, while the radial fusion is governed by diffusion.  
Thus, it is inferred that the neck degree of freedom is in the thermal equilibrium during the fusion.

Next, in order to know how the distribution reaches the equilibrium, we try to obtain a time-dependent distribution function of the neck, starting from the delta-function at $ \epsilon_0 = 1.0 $, i.e., at the initial DNC.  The Smoluchowski equation is solved, since we know that the momentum space can be approximated to be in an 
equilibrium, due to a very small inertia mass.\cite{AARS}  Then, with a linear potential, the equation to be solved is as follows,
\begin{equation}
\frac{\partial N}{\partial t}= D \frac{\partial^2 N}{\partial \epsilon^2}+ C \frac{\partial N}{\partial\epsilon},
\end{equation}
where the diffusion coefficient is $ D = T/ \gamma $, and the drift one $C = f/ \gamma $,  $ f $ being the slope parameter calculated with LDM\cite{TCP}: $ V( \epsilon) = f \epsilon $. The friction coefficient $ \gamma $ is calculated with the usual one body model.   For simplicity, we take the range of the variable $ \epsilon $ to be $ [0.0, \infty] $, instead of the realistic $ [0.0, 1.0] $ (in this case, a little more complicated expression has been obtained, but the results are essentially the same as the present case.\cite{BSAG}).  The boundary condition at $ \epsilon =0.0 $ is {\it reflective}.  With the initial and the above boundary conditions, the solution is obtained as follows,\cite{SML}
\begin{eqnarray}
N(\epsilon,t)&=& \frac1{\sqrt{4\pi Dt}}\left[\exp\left(-\frac{(\epsilon-\epsilon_0)^2}{4Dt}\right)+
\exp\left(-\frac{(\epsilon+\epsilon_0)^2}{4Dt}\right)\right]\\
&& \times \exp\left(-\frac{C}{2D}(\epsilon-\epsilon_0)-\frac{C^2t}{4D}\right)\\
&& + \frac{C}{2D} \exp\left(-\frac{C \epsilon}D\right)\cdot {\rm erfc}
\left(\frac{\epsilon+\epsilon_0-Ct}{2\sqrt{Dt}}\right).
\end{eqnarray}
For long times,  this expression becomes a Boltzmann distribution.  In Fig. 3, the time dependence is shown by distributions at various times after the contact, for the case of $ ^{100} $Mo+$^{100}$Mo system.

\begin{figure}[th]
\centerline{\psfig{file=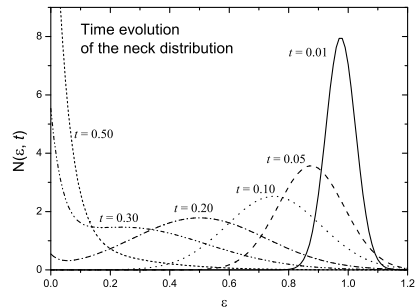,width=5.4cm}}
\caption{Time evolution of the neck distribution function is shown with examples at various times for $ ^{100} $ Mo+$ ^{100} $Mo system, for which typical values of the parameters are $D=T/\gamma$=1/8 and $C=f/\gamma$=20/8 in the unit of  MeV/$\hbar$.  The time unit is $\hbar$/MeV.}
\end{figure}

Apparently, the Boltzmann distribution in the coordinate space is established at several tenths of $\hbar$/MeV.  It is worth noticing that this time scale is far shorter than that of the 1-dimensional radial fusion discussed in the previous section.  This means that before the radial motion for fusion starts, the neck cleft is filled in, i.e., the initial DNC becomes a superdeformed mono-nucleus. During fusioning motion, we can approximately take $ \epsilon $ to be 0.0, because it is the most probable value of the Boltzmann distribution obtained above.

A similar analysis has been made for the mass-asymmetry,\cite{BSAG} which turned out that the time scale is the same order with that of the radial fusion, and thus, the two degrees of freedom have to be solved in a coupled way.\cite{2DL}

\section*{Acknowledgements}
The present work has been supported by JSPS grant No.18540268.  
One of the author (C.S.) thanks the supports from NSF of China and from NSF 
of Zhejiang Province under the grant Nos. 10675046 and Y605476, 
respectively. The authors also acknowledge supports by RCNP, Osaka Univ., 
GANIL, Huzhou Teachers College, and IPT, CEA-Saclay, which enable us to continue the collaboration.

\end{document}